\documentstyle[epsfig]{aipproc}

\begin{document}
\title{Neutrinos from Gamma Ray Bursts}

\author{Karl Mannheim}
\address{Universit\"ats-Sternwarte\\
Geismarlandstrasse 11, D-37083 G\"ottingen, Germany\\
kmannhe@uni-sw.gwdg.de, http://www.uni-sw.gwdg.de/$^\sim$kmannhe}

%\lefthead{LEFT head}
%\righthead{RIGHT head}
\maketitle

\begin{abstract}

The observed fluxes of cosmic rays and gamma rays are used 
to infer the maximum
allowed high-energy neutrino flux allowed for Gamma Ray Bursts (GRBs),
following Reference \cite{MPR_00}.
It is shown
that if GRBs produce the ultrahigh-energy cosmic
rays, they should contribute (a) at least 10\% of the extragalactic gamma
ray background between 3~MeV and 30~GeV, contrary to their observed energy flux
which is only a minute fraction of this flux, and (b) a cumulative
neutrino flux a factor of 20 below the 
AMANDA-$\nu2000$ limit on isotropic neutrinos.  
This could have two implications, either GRBs do not produce 
the ultrahigh energy cosmic rays \cite{Stecker_00,Pugliese_00}
or that the GRBs are strongly beamed and
emit most of their power
at energies well above 100 GeV \cite{Totani_99} implausibly
increasing the energy requirements, but consistent
with the marginal detections of a few low-redshift
GRBs by MILAGRITO \cite{Atkins_00},
HEGRA-AIROBICC \cite{Padilla_98},
and the Tibet-Array \cite{Amenomori_98}. 
All crucial measurements to test the models will be available
in the next few years.  These are
measurements of (i) high-energy neutrinos with AMANDA-ICECUBE
or an enlarged ANTARES/NESTOR ocean detector,
(ii) GRB redshifts from HETE-2 follow-up studies, and
(iii) GRB spectra above 10 GeV with
low-threshold imaging air Cherenkov telescopes such as MAGIC
and the space telescopes AGILE and GLAST.

\end{abstract}

\section*{Introduction}

Gamma Ray Bursts (GRBs) have for decades persisted as a new and 
unexplained phenomenon 
in astrophysics,
prompting numerous speculations about their nature and origin
\cite{Fishman_97}.
After the discovery of X-ray and optical counterparts, the cosmological origin 
of GRBs is now beyond doubt, and the physics scenario employs the
cataclysmic release of $E\sim 10^{54}$~ergs, one  solar rest mass equivalent,
under near 
vacuum conditions.  
Quasi-thermal neutrinos at $20$~MeV
energies are the expected prime carriers
during the first stages of energy release \cite{Kumar_99}, and the difficulties
in converting this energy into a high-entropy fireball during
a neutron star merger
have been pointed out 
\cite{Ruffert_98}.  The difficulties could be overcome
by assuming a disk phase with Poynting flux extraction during
coalescence\cite{Meszaros_97}, or a 
hypernova scenario \cite{Paczynski_98}, and possibly by considering
the interaction between the neutrino magnetic moment
(which arises in any model explaining finite neutrino masses) and
the magnetic field gradients in the collapsing
object.  
Owing to the
large distances of GRBs and the strong stellar background at the nuclear
binding energy scale, there is little hope for detecting the initial
neutrino emission component.
There could be a very powerful neutrino emission  component in the GeV range
due to internal dissipation driven by
proton-neutron diffusion 
\cite{Derishev_99,Meszaros_00}, and the background at these energies
is significantly lower.\\

At high energies, neutrino detection becomes easier, since the
weak interaction cross section, the muon range, and the angular
resolution all increase with energy.
As the consequence of a low mass-load (high entropy) in GRBs, 
the exploding matter can attain
ultra-relativistic velocities,
typically corresponding to Lorentz factors
$\Gamma\sim 300$, and can form relativistic shocks which are efficient
in accelerating particles to even higher energies \cite{Meszaros_92}.
This is known to occur
in the relativistic jets expelled from
Active Galactic Nuclei (AGN) \cite{Rees_67},
which transport some $E\sim 10^{60}$~ergs over their lifetime and have
typical Lorentz factors of $\Gamma\sim 10$.  As a matter of fact, 
GRB and AGN jets involve almost identical concepts in their
theoretical modeling.\\

It is thus not surprising that after the hypothesis of the
association of GRBs and ultrahigh-energy cosmic rays (UHECRs) 
by Milgrom and Usov \cite{Milgrom_95} an effort was
launched to parallel
the advanced theory of accelerated protons in
AGN 
\cite{Protheroe_83,Biermann_87,Sikora_87,Mannheim_89,Stecker_91,Mannheim_93,Rachen_93,Mannheim_95,Mannheim_98,Rachen_98} 
for GRBs 
\cite{Waxman_95,Vietri_95,Waxman_96,Waxman_97,Vietri_98,Totani_98}. 
If the proton acceleration
mechanism operates at the gyro time scale or faster, both AGN and
GRBs represent
possible classes of sources for UHECRs. 
The difference is that the AGN
produce roughly two orders of magnitude more cumulative 
flux in the observed
gamma ray band, and thus are likely much stronger sources of cosmic
rays and neutrinos, too. 
Moreover, spatial coincidences between GRBs and UHECRs
and the equality of local GRB and UHECR emissivities appear
as a poor heuristic basis for the claim of a common origin in view of the fact
that GRBs are typically too distant to allow for cosmic rays to reach Earth
unimpeded by energy losses in interactions with the microwave background. \\

While this can be regarded as an argument disfavoring the hypothesis
\cite{MPR_00,Stecker_00,Pugliese_00}, there is one
more aspect which deserves a closer look.  This aspect is
related with the shape of the multi-wavelength spectra of
relativistic jet sources.
The canonical spectral shape consists of two components,
one peaking at lower energies, and one peaking at higher energies.
The first component has a flux density that is independent
on the photon energy $E$ (in the radio/mm range in AGN
and in the X/$\gamma$-ray range in GRBs) which steepens by the
factor $E^{-1}$
at higher energies.  This part of the spectrum is successfully
modeled as the synchrotron radiation from accelerated
electrons. If protons are accelerated in
the sources, too, then there must be a second peak due
to the hadronically-induced radiation at high energies in the canonical
spectrum.  Such a component
has been found in the spectra of AGN jets in the MeV-to-TeV range,
albeit it could also be interpreted as being due to inverse-Compton
scattering of low-energy photons by the synchrotron-emitting electrons
\cite{Mannheim_98}.
Naturalness would then provoke the assumption that GRB spectra have
a corresponding peak 
at ultrahigh energies which has escaped detection so far.\\

This is consistent with the fact that due to the typical high redshifts
of GRBs, such an emission component would be attenuated by pair production
in traversing the intergalactic radiation fields \cite{Stecker_98,Kneiske_00}. 
We could be fooled by beaming statistics, if
the dominant isotropic
gamma ray output of GRBs went into a diffuse channel, e.g. due
to cascading in the intergalactic medium wiping out directional 
memory and temporal correlation
\cite{Totani_98}.\\

The issue highlights the importance of neutrino observations. Neutrinos
suffer practically no energy losses (except for adiabatic losses owing
to the expansion of the Universe) and can therefore be used to measure
the flux of cosmic rays escaping from GRBs.
Prompt UHE
protons escaping from relativistic
outflows suffer from synchrotron and adiabatic losses 
\cite{Rachen_98}, and therefore
the neutron-escape fraction of
cosmic rays from internal shocks should be the dominant one.
Protons from external shocks would have a too steep spectrum
to be important at UHE energies for reasonable energetics \cite{Kirk_00}.
The neutron-fraction of the escaping cosmic ray flux can be inferred from 
measurements of neutrinos in the 100~TeV range.  These neutrinos would
be correlated with the electromagnetic bursts.
Cosmic rays also give rise to an uncorrelated secondary neutrino flux
during propagation through the microwave background
due to photo-production of pions.\\

The paper first provides
a general bound on the neutrino emission from GRB-like sources,
constrained by the observed diffuse gamma-ray and
cosmic ray proton fluxes.  Secondly, more specific assumptions
regarding the spectral shape of the emitted protons 
are employed to find out the expected flux
of high-energy neutrinos following the Milgrom \& Usov hypothesis.
Finally, this would have consequences for the properties of high
energy gamma-ray emission from GRB which are briefly discussed.

\section*{Upper limit for GRB-like photo-hadronic
sources of neutrinos}

Recently, Mannheim, Protheroe, and Rachen  \cite{MPR_00} (MPR) have
derived an upper limit for the neutrino flux from extragalactic
sources, in the framework of photo-production models in which the
neutrinos originate from pions produced in hadronically-induced
interactions of accelerated protons with low-energy photons.
With every neutrino, there is a minimum number of photons and 
neutrons associated, and the flux of those is not allowed to exceed
observed omnidirectional fluxes of cosmic ray protons and gamma rays.
In their Fig.~3, MPR show this limit for sources 
with conditions typical for AGN (steady sources with
flux density spectral index
$\alpha=1$).
Here, the focus is on the short-lived GRBs in 
which the synchrotron target photons have
a flat spectrum ($\alpha=0$).
The relation between cosmic ray and neutrino fluxes at ultrahigh-energies
(above $\sim 10^7$~GeV proton energy in the comoving frame)
is then given by 
\begin{equation}
\label{qnu}
Q_{\nu_\mu}(E)=416 Q_{\rm cr}(25E)
\end{equation}
i.e. the neutrino yield can
be up to factor of 5 larger than in the case of AGN \cite{Muecke_99}.\\

The experimental upper limit on a possible extragalactic proton
contribution to the local flux of cosmic rays is consistent
with an extrapolation of the
observed, presumably protonic, spectrum at
$10^{10}$~GeV down to $10^6$~GeV using a spectral index of 2.75.
Owing to the steepness of this spectrum, saturating the upper limit
implies an increasing emissivity of the putative GRB-like
sources with decreasing energy.  The electromagnetic and neutrino
luminosities of GRB-like sources are the same.
Below PeV energies, their electromagnetic output 
begins to reach the level of the observed extragalactic gamma ray
background and the upper limit levels off.
Generally, since gamma-rays from pion decays and from
pair production are reprocessed by electromagnetic cascades,
only the bolometric (integrated) fluxes can be compared.  \\

\begin{figure}[b!] % fig 1
\centerline{\epsfig{file=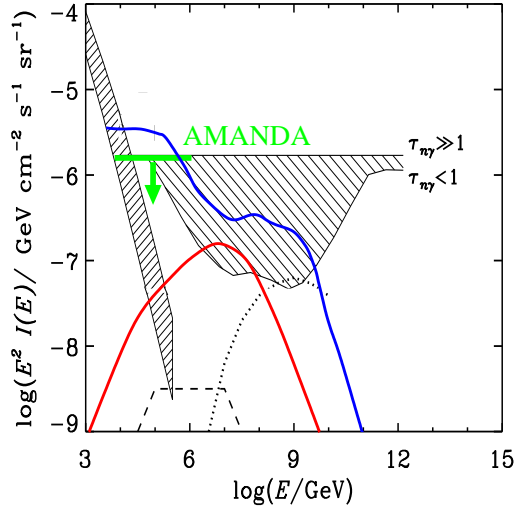,height=3.0in,width=3.5in}}
\vspace{10pt}
\caption{Sketch of the neutrino upper limit ($\nu_\mu+\bar\nu_\mu$) 
for GRB-like
photo-hadronic sources ({\it upper thick solid line})
and generic GRB flux if Milgrom \& Usov
hypothesis were true ({\it lower thick solid line}).
For comparison, the upper limit for AGN-like sources
is shown in the optically thin
({\it thin line labeled $\tau_{n\gamma}<1$}) and optically thick 
({\it thin line labeled $\tau_{n\gamma}\gg 1$}) cases, respectively.
Also shown is the atmospheric neutrino flux between horizontal
and vertical directions ({\it hatched narrow band}), and the
experimental limit from AMANDA (as reported at
the $\nu$-2000 conference by S.~Barwick). For
further details, see text.} 
\label{bound}
\end{figure}

Equation (\ref{qnu}) still ignores the effect of cosmic ray attenuation, which
is important for sources with strong redshift evolution.  
A strong redshift
dependence $\propto (1+z)^{3.4}$ of the cumulative
source emissivity is assumed for the GRBs.  This corresponds
to the observed dependence for star formation in galaxies and AGN
\cite{Boyle_98}.
Above $10^{10}$~GeV, the upper limit spectrum steepens rapidly
due to the inescapable energy losses of pions and muons in
compact, variable sources
\cite{Rachen_98}. Below
PeV energies, the increase is only by
a factor of 2, since 
the constraint from the diffuse gamma-ray background imposed
by $L_\gamma\simeq L_\nu$ (where $L_i$ denotes the bolometric luminosity
emitted in species $i$) becomes the stronger one,
and AGN-like conditions have $L_\gamma=2L_\nu$ (due to the
contribution of gamma-rays from Bethe-Heitler pairs).
For realistic model assumptions, the turnover at $10^{10}$~GeV
would occur at even lower energies (see next section).
Figure\,\ref{bound} also shows previously predicted fluxes
for GRBs based on a simplified treatment
of the photo-production kinematics
\cite{Waxman_97} and for AGN \cite{Mannheim_95}
for comparison.\\

\section*{Generic GRB neutrino flux}
For the calculation of an upper limit
for GRB-like sources, explicit model assumptions
required for a generic GRB model have not been made.
A crucial assumption is that of the slope of the
energy distribution of accelerated protons.\\

The external shock where the ejecta impact the external
medium is highly relativistic and should thus produce
particle spectra $\propto E^{-2.2}$ \cite{Kirk_00}.
If protons accelerated at the external shock were to
explain the UHE cosmic rays, the power in the low-energy
part of the population would have to be unrealistically high,
and the GRBs radiatively extremely inefficient.\\

Internal GRB shocks are expected to be only mildly relativistic
owing to the velocity differences of superseding shells being
smaller than those with the external medium, and therefore
a proton spectrum $\propto E^{-2}$ seems possible.  Such an energy
distribution
delivers a large fraction of the energy in the proton population
at UHE energies.  The shape of the neutrino spectrum follows
$\propto E^{-1}$ below $E_{\rm b}=100 (\Gamma/300)^2$~TeV and
$\propto E^{-2}$ above, until the spectrum steepens above 10~PeV
due to energy losses of the muons in the expanding ejecta.
The actual spectrum will be slightly more curved (see
the lower thick solid curve in Fig.~\ref{bound}), since the
photo-hadronic
interactions are dominated by the $\Delta$-resonance
in the 100 TeV range, 
whereas by multiple pion production above PeV lab frame energies.
A more accurate treatment using the 
SOPHIA code \cite{Muecke_99} must await future work.\\

Since the protons must suffer synchrotron and adiabatic losses upon escape
\cite{Rachen_98},
the neutrons produced
in photon collisions dominate the escaping fraction of cosmic
ray baryons at ultrahigh energies.
The generic GRB neutrino flux following from the assumption
that GRBs produce the UHE cosmic rays 
above $10^9$~GeV (exempting the super-Greisen events), 
is shown in Fig.\,\ref{bound}.
The corresponding integral gamma-ray flux is given by 
$F_\gamma=
{3\over 2}(2\times 10^{-7}\times \ln[100])$~GeV~cm$^{-2}$~s$^{-1}$~sr$^{-1}$
which corresponds to 10\% of the extragalactic gamma ray background
between 3~MeV and 30~GeV ($1.5\times 10^{-5}$ in the same units).
Comparing with the extragalactic gamma ray background between
1~GeV and 30~GeV
($4.2\times 10^{-6}$),
the number is 33\%.
The factor $\ln[100]$ takes into account the bandwidth of the
neutrino spectrum (the limit for AGN-like sources shown in Fig.~\ref{bound}
takes into account a factor $\exp[1]$ as the bolometric correction
for trial spectra of the form $\propto E^{-1}\exp[-E/E_{\rm max}]$).\\

Note, however, that the secondaries produced during
propagation of UHE cosmic rays in interactions with the microwave
background add another (albeit uncorrelated, and therefore
truly diffuse) neutrino component at EeV energies 
(similar to the dotted curve in Fig.~\ref{bound}) and
also add an electromagnetic component emerging in the observed gamma-ray
band.  This component has an energy flux level comparable to the
minimum of the upper limit for AGN-like sources at $10^9$~GeV,
since photo-production
occurs close to the threshold energy in interactions with photons from the
microwave background. 
The spectrum is rather sharply peaked at UHE energies and has a 
bolometric correction of
$\sim\exp[1]$.
The corresponding gamma-ray flux is 
$2\times 10^{-7}$~GeV~cm$^{-2}$~s$^{-1}$~sr$^{-1}$, 
so that the combined GRB contribution to the extragalactic
gamma ray background would be $1.6\times 10^{-6}$ 
in the same units (11\% of the 3~MeV-30~GeV
background or 38\% of the 1~GeV-30~GeV background).\\

Constraints on the total isotropic energy per burst, their radiative
efficiency, and
the burst rate have not been imposed here.  It is clear, however,
that if GRBs shall explain the UHE cosmic rays, the current physical
interpretation of GRBs is severely challenged by the 
dramatically increased energy demands.

\section*{Implications for high-energy gamma-ray tails}
The powerful high-energy gamma-ray emission implied by the
Milgrom \& Usov hypothesis raises the possibility for
falsification by gamma-ray observations.  However, 
the current generation
of ground-based telescopes are sensitive only above  300~GeV
corresponding to a pair-attenuation length of $z=0.3$.
Only a small number of bursts is expected to occur at redshifts
less than 0.3.  From an analysis of the 2nd BATSE catalogue data,
the expected rate of 100 per year\cite{Mannheim_96} is too low for the
small field of view of imaging air Cherenkov telescopes
to make a chance detection.  The large-field-of-view detectors
MILAGRITO, Tibet, and HEGRA-AIROBICC have not found 
strong evidence for GRBs,
but their thresholds are in the TeV domain where less than
one GRB per year is expected to occur.  It is interesting to
note, however, that the groups have all reported
marginal evidence for single GRBs which is statistically 
conceivable
\cite{Atkins_00,Padilla_98,Amenomori_98}.
The implied flux is
consistent with the Milgrom \& Usov hypothesis \cite{Totani_99}.
This high-energy tail
would have an extremely hard spectrum rising 
above 10 GeV, since EGRET has not discovered
evidence for an upturn.  Clearly, this is a challenging 
possibility to be tested by AGILE, GLAST, or MAGIC.

\section*{Summary and conclusions}
Non-thermal power law emission and afterglows over a broad
range of wavelengths indicate the acceleration of charged particles to
relativistic energies in GRBs.  
If GRBs are the sources of UHE cosmic
rays, they must produce at least 10\% of the extragalactic
gamma ray background between 3~MeV and 30~GeV and a diffuse flux
of neutrinos in the 100~TeV to 10~PeV regime which is a 
factor of 20 below the AMANDA-$\nu$2000 limit on 
isotropic neutrinos.  The time stamp associated with GRBs
increases the sensitivity further compared with the sensitivity for
an isotropic, uncorrelated background.
Low-redshift GRBs would also have powerful high-energy tails in their 
gamma ray spectra 
which are consistent with the marginal detections reported
by the MILAGRO, HEGRA, and Tibet Collaborations.
The next years will bring rapid progress to the understanding of GRBs
and their high-energy emissions, 
with improved measurements of GRB neutrinos, redshifts,
and high-energy gamma-ray tails using AMANDA-ICECUBE/ANTARES/NESTOR,
HETE-2 follow-up programs, and MAGIC/AGILE/GLAST.

\acknowledgments
I am deeply indebted to J\"org Rachen and Raymond Protheroe for
their invaluable contributions to this research. 
Superior support through a Heisenberg-Fellowship granted
by the DFG is gratefully acknowledged.

\end{document}